\documentclass[sigconf]{acmart}

\usepackage{comment}
\usepackage[np, autolanguage]{numprint}
\usepackage{paralist}
\usepackage{enumitem}
\usepackage{xcolor}

\newcommand{\Hypothesis}[2]{
    \begin{description}[noitemsep,itemsep=-2pt,nosep,topsep=-1pt,leftmargin=*]
    \phantomsection\label{section:setup:hypothesis#1}
    \item[H#1] #2
    \end{description}
    \medskip
}

\newcommand{\HypothesisRef}[1]{\textbf{\hyperref[section:setup:hypothesis#1]{H#1}}}

\AtBeginDocument{%
  \providecommand\BibTeX{{%
    \normalfont B\kern-0.5em{\scshape i\kern-0.25em b}\kern-0.8em\TeX}}}

\copyrightyear{2020} 
\acmYear{2020} 
\setcopyright{acmlicensed}\acmConference[UMAP '20]{Proceedings of the 28th ACM Conference on User Modeling, Adaptation and Personalization}{July 14--17, 2020}{Genoa, Italy}
\acmBooktitle{Proceedings of the 28th ACM Conference on User Modeling, Adaptation and Personalization (UMAP '20), July 14--17, 2020, Genoa, Italy}
\acmPrice{15.00}
\acmDOI{10.1145/3340631.3394862}
\acmISBN{978-1-4503-6861-2/20/07}

\begin{document}
\fancyhead{}
\title{Eliciting User Preferences for Personalized Explanations for Video Summaries}

\author{Oana Inel}
\affiliation{%
  \institution{Delft University of Technology}
     \city{Delft}
    \country{the Netherlands}
}
\email{o.inel@tudelft.nl}

\author{Nava Tintarev}
\affiliation{%
  \institution{Delft University of Technology}
     \city{Delft}
        \country{the Netherlands}
}
\email{n.tintarev@tudelft.nl}

\author{Lora Aroyo}
\authornote{This author has done the work in the capacity of a PhD advisor for Oana Inel}
\affiliation{%
  \institution{Google}
     \city{New York}
        \country{USA}
}
\email{l.m.aroyo@gmail.com}
\renewcommand{\shortauthors}{Inel et al.}

\begin{abstract}
Video summaries or highlights are a compelling alternative for exploring and contextualizing unprecedented amounts of video material. However, the summarization process is commonly automatic, non-transparent and potentially biased towards particular aspects depicted in the original video. Therefore, our aim is to help users like archivists or collection managers to quickly understand which summaries are the most representative for an original video. In this paper, we present empirical results on the utility of different types of \emph{visual explanations} to achieve transparency for end users on how representative video summaries are, with respect to the original video. We consider four types of video summary explanations, which use in different ways the concepts extracted from the original video subtitles and the video stream, and their prominence. The explanations are generated to meet target \emph{user preferences} and express different dimensions of transparency: \emph{concept prominence}, \emph{semantic coverage}, \emph{distance} and \emph{quantity of coverage}. In two user studies we evaluate the utility of the visual explanations for achieving transparency for end users. Our results show that explanations representing all of the dimensions have the highest utility for transparency, and consequently, for understanding the representativeness of video summaries.

\end{abstract}




\maketitle


\section{Introduction}
\label{sec:introduction}

Online videos constitute the largest, continuously growing portion of Web content, reaching a daily upload of 500 hours of video\footnote{\url{https://bit.ly/2Guh3Gh}, June 2019}. A recent forecast suggests that by 2022 video traffic will be 82\% of all Web traffic \cite{cisco2018cisco}. There is also an increased interest in watching videos online, with one billion hours of video being watched every day, only on YouTube\footnote{\url{https://youtube.
googleblog.com/2017/02/you-know-whats-cool-billion-hours.html}}. However, despite effortless access to a large number of videos, it is not trivial for lay users or even professional users like journalists, archivists, or collection managers, to meaningfully consume all this video information. Typically, to explore a topic they need to watch an overwhelming amount of video material. To help processing this amount of video material, \textit{video summaries} or \textit{video highlights}, have been introduced.

There is a multitude of automated video summarization techniques \cite{song2015tvsum,zhang2016video,vasudevan2017query,zhou2018deep} used to generate light-weight previews of long video materials.
To generate video summaries, video screenshots or fragments are usually chosen based on how well they \textit{represent} the video \cite{chakraborty2015adaptive,Kanehira_2018_CVPR,liu2002optimization}. 
This automated process, however, is prone to amplify or diminish certain aspects of the video, it might omit meaningful information and thus, might lead to the \textit{misrepresentation} of the original content. For example, consider watching a news broadcast announcing a powerful storm. A video summary could show only the name of the last area hit by the storm, and a random day in which it is expected to reach an area. Thus, the decision of what is representative is an automated process, that lacks human oversight and can be misleading due to such automation~\cite{diakopoulos2017algorithmic}. 

In this paper, we advocate for a novel user-centric solution to provide a quick and easy decision making support on the representativeness of video summaries with regard to the original video. Our goal is to empower users with an efficient and effective visual decision support that helps them to quickly understand the differences in content of the original video and its summary. Therefore, we use \emph{user preferences} to model \emph{visual explanations} along various \emph{dimensions of transparency} (\emph{i.e.}, \emph{semantic coverage}, \emph{semantic prominence}, \emph{quantity coverage} and \emph{distance}), to increase people's awareness with regard to: \emph{(1)} the concepts that are present in the original video and \emph{(2)} how well these concepts are \textit{represented} in the video summary compared to the original video. Our novel approach for generating video summary explanations combines the output of video analysis tools, such as key concepts (\emph{i.e.}, people, events, organizations, locations and other concepts represented in the videos \cite{gligorov2011role}, \cite{lai2014video}) from video subtitles (named entity extraction) and video streams (video labeling)
and represents them in two types of visual explanations, namely, word clouds and donuts, along the \emph{four dimension of transparency}. 

The proposed solution to generate visual explanations is generalizable to both short and longer videos, provided that only key concepts are depicted in the visual explanations. The length of the video summaries does not affect the efficiency of generating such visual explanations, given that, in general, video summaries should be short and concise. Answering the research question: \textit{What kind of visual explanations are most useful to understand how representative a video summary is with regard to the original video?}, we make the following contributions:

\begin{itemize}[noitemsep,nosep,topsep=0pt,leftmargin=*]
    \item a set of \emph{user preferences} for personalizing visual explanations for video summaries;
    \item an \emph{annotated corpus of \np{200} videos with key concepts} extracted from video subtitles and video frames;
    \item an \emph{annotated corpus of \np{800} video summaries (four summaries per video) with concepts} extracted from video subtitles and frames;
    \item a \emph{corpus of \np{3200} visual explanations} (four explanations per video summary, with different levels of \emph{transparency});
    \item results from \emph{two user studies} evaluating the \emph{utility of four types of visual explanations} which are modeled based on \emph{user preferences} and along \emph{four dimensions of transparency} to understand the \emph{representativeness} of video summaries;
\end{itemize}

Our results show that users find most useful for understanding the representativeness of video summaries the explanations with the highest level of transparency - combining all four dimensions (\emph{i.e.}, show the amount and the prominence of topics covered and not covered in the video summary, compared to the original video).  All data and code are publicly available on GitHub\footnote{The scripts to replicate the approach, the results of the user studies and their analysis, as well as the dataset (videos, video summaries and visual explanations) are available at \url{https://github.com/oana-inel/FAIRView-VideoSummaryExplanations/}}.


\section{Related Work}
\label{sec:related_work}

Although the focus of this paper is not on generating video summaries, we consider the overview of work in the area of video summarization useful to understand the potential issues of automatically generated summaries. We also review approaches for creating personalized video summary explanations. Finally, we discuss limitations and opportunities.

\subsection{Video Summarization Approaches}
\label{subsec:rw_approaches}

Current literature identifies three types of video summaries: static \cite{de2011vsumm,mahasseni2017unsupervised,vasudevan2017query} composed of keyframes, dynamic \cite{wu2011video,jin2017elasticplay} composed of keyshots and hybrid \cite{DBLP:conf/visapp/AinasojaHLK18}. Video summaries can be created using techniques that use a query \cite{sharghi2016query,vasudevan2017query}, or without a query \cite{DBLP:conf/visapp/AinasojaHLK18,zhang2016video}. 

Supervised methods \cite{gong2014diverse,zhang2016video,vasudevan2017query} need manually created video summaries to learn the features of the keyframes and keyshots to be included in the summary. Ground truth datasets are either created automatically \cite{de2011vsumm} or crowdsourced \cite{gong2014diverse,vasudevan2017query,Kanehira_2018_CVPR}. In contrast, unsupervised video summarization \cite{mahasseni2017unsupervised,DBLP:conf/airs/InoueY18,zhou2018deep,DBLP:conf/visapp/AinasojaHLK18} relies on pre-defined criteria to select the keyframes and keyshots of the summary. Common criteria include color features and clustering \cite{de2011vsumm,jeong2015consumer}, interestingness \cite{gygli2014creating}, importance and relevance \cite{sharghi2016query,sharghi2017query}, representativeness and uniqueness \cite{chakraborty2015adaptive}, diversity and representativeness \cite{Kanehira_2018_CVPR,liu2002optimization,vasudevan2017query}, among other. This all leads to the fact that multiple summaries can be created for the same video~\cite{sharghi2016query,sharghi2017query,Kanehira_2018_CVPR}.

Benchmarks such as TVsum \cite{song2015tvsum} and SumMe \cite{gygli2014creating} are commonly used for qualitative evaluation of video summaries, where performance is reported in terms of F1-score, mean average precision, among others. User studies are typically used for qualitative evaluation. Video summaries are evaluated with various likert scales on their quality \cite{Kanehira_2018_CVPR}, informativeness and enjoyability \cite{DBLP:conf/airs/InoueY18}, or how much a summary helped to make sense of the full video \cite{wu2011video}.

\subsection{Personalization of Video Summaries}
\label{subsec:rw_explanations}

Video summary personalization focuses on generating video summaries given a user's query \cite{sharghi2017query,sharghi2016query,vasudevan2017query}. \citet{ghinea2014novel} proposed a summarization algorithm that creates a user profile, \emph{i.e.}, the user sees a set of 25 concepts present in the video and indicates through a list or a sliding window which or how much of these concepts should be included in the summary. \citet{jin2017elasticplay} integrated fast-forward functionality, allowing users to skip parts of the summaries, while \citet{chongtay2018responsive} introduced the idea of responsive news summarization, \emph{i.e.}, automatically creating news summaries of different lengths, while providing access to the full news item.

Explanations have been extensively developed in the context of recommender systems \cite{tintarev2012evaluating}, but little research is found in the context of video summarization. The ANSES system \cite{pickering2003anses} can be seen as a first summary personalization attempt. It creates news broadcast summaries by enriching video subtitles with named entities of type organization, person, location and date, and showing them to users for a quick overview of the summary content.   

\subsection{Limitations and Opportunities}
\label{subsec:rw_limitations}

Research on video summarization is extensively focused on increasing the accuracy of the automated summarization tools, but it lacks focus on transparency. Users are faced with condensed video summaries without understanding the underlying decisions taken to generate the summary. Previous work \cite{jin2017elasticplay} also suggests that users prefer more \emph{transparent} video summarization, and that they express concerns regarding the black box process of automated skipping through video frames or shots. This suggests an opportunity to study how to best provide transparency for video summarization. Moreover, even though \emph{representativeness} is a fundamental criteria to generate video summaries, they are rarely evaluated in terms of how well they are perceived to represent the original video by end users. Thus, the aspect of representativeness can potentially be included in explanations accompanying the video summaries.

In this work we address these limitations and focus on increasing user awareness of the representativeness of the video summary with regard to the original video, through \emph{visual explanations with different dimensions of transparency}. We aim to empower end-users of video summaries, such as media researchers and video archivists, with tools that can help them evaluate the representativeness of video summaries with regard to the original video. To this end, we investigate a novel approach, which makes use of concepts overview, \emph{e.g.}, people, organizations, events, locations, and other concepts, to create visual explanations for more transparent video summaries. Thus, instead of explaining how the video summary was generated, we focus on empowering users with a personalized, user-centric visual description that they can use to make better choices for which summaries represent best an original video.


\section{Video Dataset}
\label{sec:dataset}

We selected a random subset of 200 short (1-2 minutes, with an average duration of 99 seconds) news videos from a publicly available dataset of English-language videos \cite{jong2018human}. We chose short videos because: \emph{(1)} people's attention span is constantly declining (\emph{i.e.}, \cite{lopatecki2019passively} reports a decrease in people attention span after watching a video for 75 seconds); and \emph{(2)} shorter videos are more suitable for user studies both in terms of raters attention and with respect to the overall cost. We used Speech Transcription offered by the Google Video Intelligence API\footnote{\url{https://cloud.google.com/video-intelligence/docs/transcription}, retrieved August '19} to extract all video subtitles.

\paragraph{Video Summaries}
\label{sec:dataset_video_summaries}
We generated four video summaries for each video in our dataset using an off-the-shelf video summarization approach \cite{markatopoulou2018implicit} focusing on news videos and preserving video chronology. For each video, we varied the length of the summaries. Thus, for each video we have two summaries of 10 seconds and two summaries of 20 seconds. A video summary is composed of video segments of three to six seconds, which are concatenated, until reaching the desired length (\emph{i.e.}, 10 or 20 seconds). All video segments included in the summary follow the original timeline of the video and all video summaries contain the audio track to accommodate the user preferences \textbf{UP1} and \textbf{UP2} in Section \ref{sec:req_user_study} (\emph{i.e.}, preserve video segments chronology and preserve audio track).


\section{Video Summary Explanations}
\label{sec:proposed_model}

In this section we present our methodology for modeling and creating visual explanations, which aim to improve video transparency and help users assess the representativeness of a summary. We first describe a qualitative user study that we conducted to gather \emph{user preferences} for personalizing video summaries in Section \ref{sec:req_user_study}. Based on these user preferences we derive \emph{four dimensions of transparency} and design \emph{four types of visual explanations} (Section \ref{sec:types_of_explanations}). In Section \ref{sec:methodology_of_explanation_generation} we present our methodology for generating such explanations.

\subsection{User Preferences Elicitation}
\label{sec:req_user_study}

We conducted a small scale user survey, with four media professionals - media researchers, audio-visual collection owners and archivists - to gather insights on \emph{(1)} the characteristics of useful video summaries and \emph{(2)} effective interaction with video summaries by means of explanations. The user survey was based on the literature review in Section \ref{sec:related_work} and the observations in Section \ref{subsec:rw_limitations}. Below, we present the \emph{user preferences} (\textbf{UP}) derived from this study. 

Participants prefer a video summary in which \emph{the chronology of the scenes is preserved} (\textbf{UP1}), since chronology is strongly correlated with understanding the narrative of a viewpoint. Furthermore, they also prefer a video summary accompanied by the \emph{audio track}, \emph{a link to the original video} and, optionally, \emph{subtitles} and \emph{a textual summary of the video}, as part of \textbf{UP2}. 

We were also interested to understand what would make a video summary explainable, by focusing on the \emph{type of explanations}, the \emph{format of the explanations}, and the \emph{visualization of explanations}. For the type of explanations, \textbf{UP3}, participants prefer to have \emph{an overview of both concepts that are covered by the video summary} (to guide an understanding of the original content) and \emph{concepts that are not covered by the video summary} (to contextualize the original video). Participants also prefer \emph{short textual explanations} (\textbf{UP4}), which can be checked after watching the video summary, or, \emph{as video subtitles}. Moreover, the \emph{explanations should be shown as a side info box, while the video summary runs}, \textbf{UP5}, so that they are not disturbed while watching the video summary. 

\subsection{Types of Video Summary Explanations}
\label{sec:types_of_explanations}

We now introduce the \emph{dimensions of transparency}, and the concrete \emph{visual explanations} we propose based on the user preferences identified in Section \ref{sec:req_user_study}. We identified four dimensions of transparency to consider when generating video summary explanations. They comply with user preference \textbf{UP3}, which emphasizes the importance of understanding the similarities and differences between summaries and original video, in terms of semantics and quantities: 
\begin{itemize}[noitemsep,nosep,topsep=0pt,leftmargin=*]
\item \emph{semantic coverage}: concepts covered in a summary and/or original video;
\item \emph{semantic prominence}: concept prominence in a summary and/or original video;
\item \emph{quantity coverage}: fraction of concepts covered and not covered in the summary;
\item \emph{distance}: difference between the summary and the original video.   
\end{itemize}

We use wordclouds and donuts charts for our visual explanations. Because users had a strong preference for textual explanations for understanding the topics covered, \textbf{UP4}, we consider wordclouds to balance well text and graphics. As we aim for a quick and easy decision making support, narratives, or natural language explanations would not be a feasible solution due to the large amount of information, \emph{i.e.}, all the concepts present in a video that need to be explained. Wordclouds are also well representing the dimensions of \emph{semantic coverage}, \emph{semantic prominence} and \emph{distance}. Donut charts offer a very quick overview of the total overlap of concepts between the original video and the summary, thus representing the \emph{quantity coverage} and \emph{distance} dimensions. We represent the visual explanations in green and purple colors because these two colors together are accessible for (the majority of) people with color blindness.

\begin{figure}[!ht]
  \centering
  \includegraphics[width=0.7\linewidth]{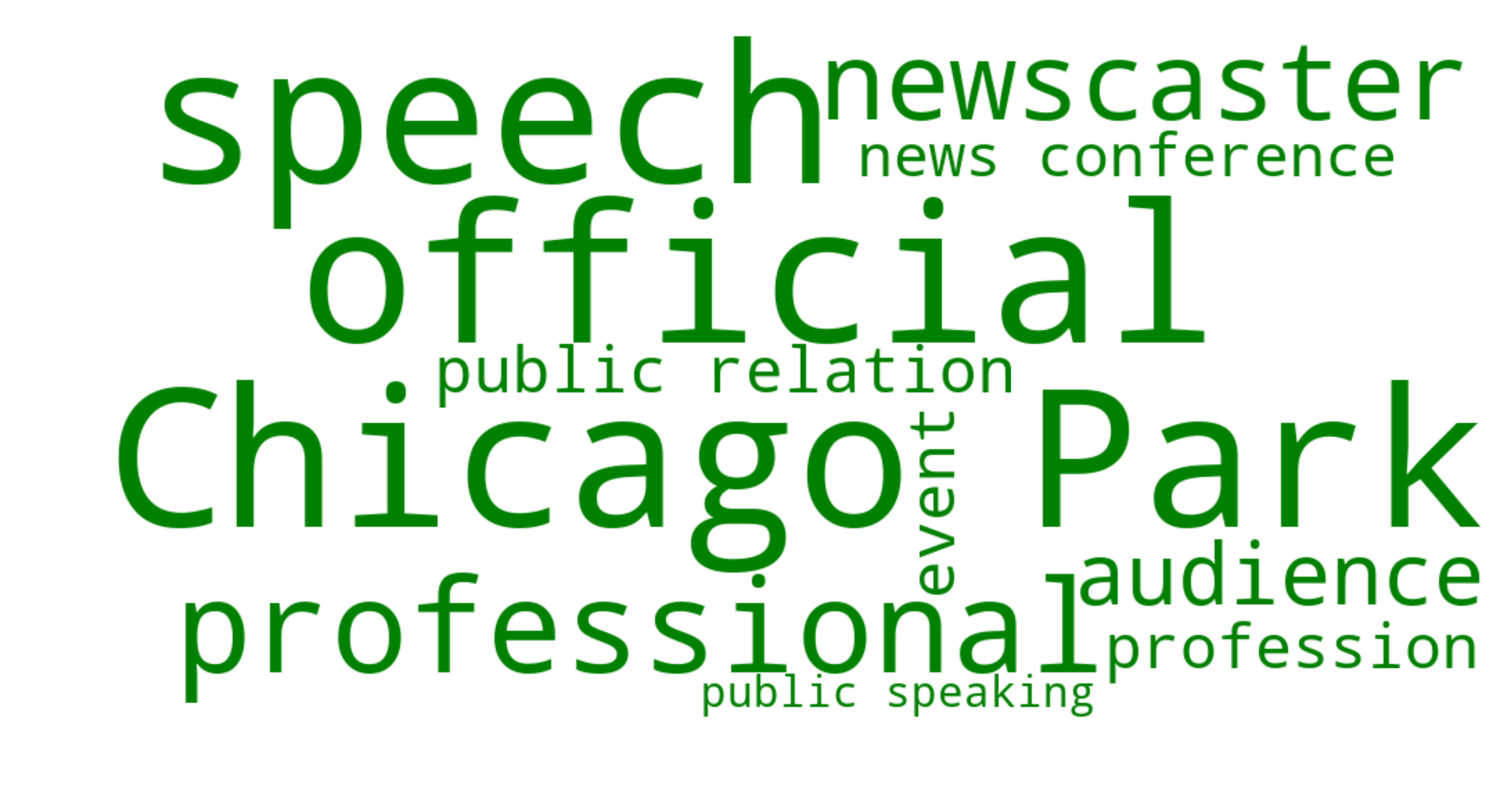}
  \caption{Summary WordCloud: The words in green are concepts found in the video summary. The word size indicates prominence of the concepts in the video summary.}
  \label{fig:visual_explanation_1}
\end{figure}

Our participants mentioned in \textbf{UP3} the need to see \emph{(1)} which are the main topics covered in the video summary -- this aspect is supported in the \texttt{Summary WordCloud} explanation and \emph{(2)} which are the main topics from the original video that are not covered by the summary -- this aspect is supported in the \texttt{Overlap WordCloud} and \texttt{Overlap Fraction} explanations, where the latter indicates the proportion of overlap between topics. Finally, the \texttt{Combined WordCloud + Fraction} explanation combines all the visualizations and all dimensions, and consequently represents both points \emph{(1)} and \emph{(2)}. We describe each visualization individually below.

\textbf{\emph{Summary WordCloud} (baseline)}: depicted in Figure \ref{fig:visual_explanation_1}, represents the \emph{semantic coverage} and \emph{semantic prominence} dimensions in the video summary. It summarizes the main concepts depicted in the video summary, in a word cloud. The concepts (words) covered by the video summary are depicted with \textbf{green} color. The size of the words indicates the prominence of the concepts, \emph{i.e.}, the larger the words, the more prominent the concept in the video summary.

\begin{figure}[h]
  \centering
  \includegraphics[width=0.7\linewidth]{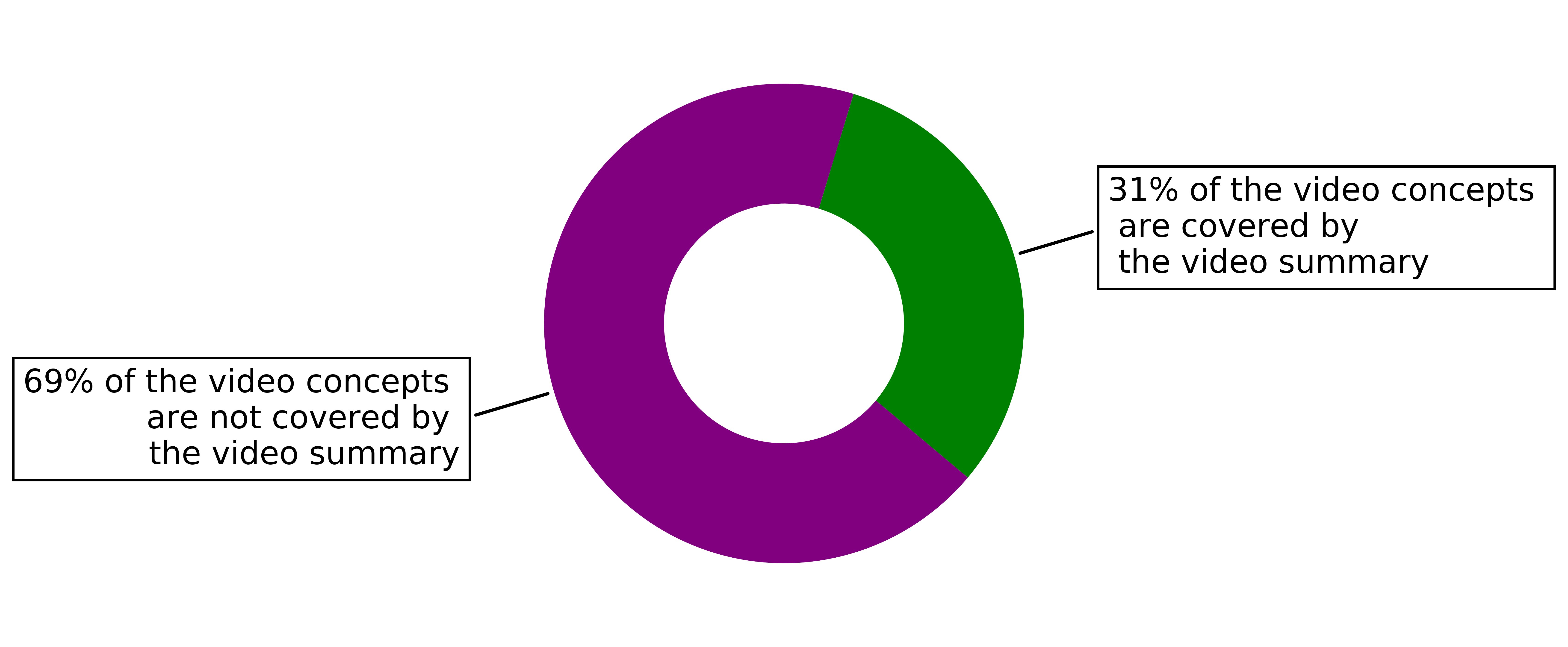}
  \caption{Overlap Fraction: The portion in green shows the percentage of original video concepts covered in the video summary. The portion in purple shows the percentage of original video concepts not covered in the video summary.}
  \label{fig:visual_explanation_2}
\end{figure}

\textbf{\emph{Overlap Fraction}}: depicted in Figure \ref{fig:visual_explanation_2}, represents the \emph{quantity coverage} and \emph{distance} dimensions. It summarizes the percentage of concepts in the original video that \emph{(1)} is covered by the video summary in \emph{green} color and that \emph{(2)} is not covered by the video summary in \textbf{purple} color, in a donut chart. 

\begin{figure}[h]
  \centering
  \includegraphics[width=0.7\linewidth]{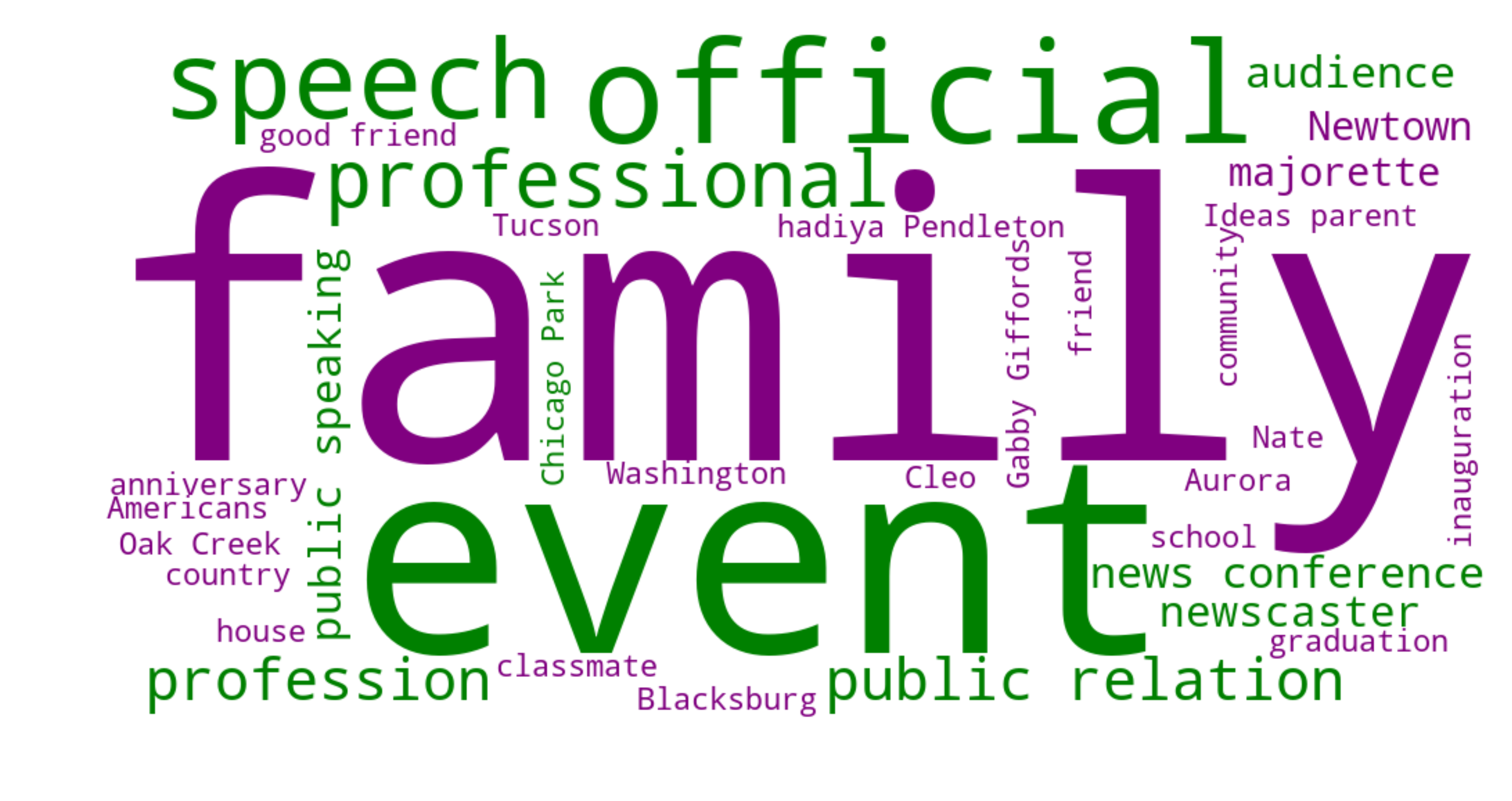}
  \caption{Overlap WordCloud: The words in green are concepts found in the video summary. The words in purple are concepts found in the original video and which are not found in the video summary. The word size indicates prominence of the concepts in the original video.}
  \label{fig:visual_explanation_3}
\end{figure}

\textbf{\emph{Overlap WordCloud}}: depicted in Figure \ref{fig:visual_explanation_3}, represents the \emph{semantic coverage}, \emph{semantic prominence} and \emph{distance} dimensions. It summarizes the main concepts depicted in the video summary in \textbf{green} and the main concepts in the original video, that are not covered by the summary in \textbf{purple}, in a word cloud. 
The size of the words indicates the prominence of the concepts, \emph{i.e.}, the larger the word, the more prominent the concept is in the original video. 

\textbf{\emph{Combined WordCloud + Fraction}}: depicted as Figure \ref{fig:visual_explanation_3} and Figure \ref{fig:visual_explanation_2} together, represents all four dimensions: \emph{semantic coverage}, \emph{semantic prominence}, \emph{quantity coverage} and \emph{distance}. It summarizes the main concepts depicted in the video summary and the main concepts that are found in the original video, but are not covered by the summary and their coverage in percentage. The visual explanation is a combination of the word cloud proposed in \emph{Overlap WordCloud} and the donut proposed in \emph{Overlap Fraction}. The colors and the size of the words have the same meaning as described above.

\subsection{Explanation Generation Methodology}
\label{sec:methodology_of_explanation_generation}

Further, we describe the methodology applied to semantically enrich the videos in our dataset, \emph{i.e.}, to identify the relevant concepts. We applied two types of machine enrichment, namely entity extraction from the video subtitles and label extraction from the video stream, to generate a list of concepts that appear in the video. 

\paragraph{Concept Extraction}
First, we extracted (common and proper) \textbf{entities} from the \emph{video subtitles}, by running the Google Natural Language API\footnote{\url{https://cloud.google.com/natural-language/}, retrieved August 2019}. Each entity may have assigned a type
, from a predefined list: person, location, organization, event, among others. Second, we extracted \textbf{labels} from the \emph{video stream} using the Google Video Intelligence API\footnote{\url{https://cloud.google.com/video-intelligence/}, retrieved August 2019}. We identify all labels for each video frame (a frame refers to a one second video fragment). Similarly, each label might have a type assigned from a predefined list.

Several concepts identified in the video subtitles and video frames could refer to the same concept (same lemma), but have different forms. Thus, to better align the concepts we first run part-of-speech (POS) tagging\footnote{NLTK POS tagger \url{https://www.nltk.org/book/ch05.html}, retrieved Sep.'19} on the concepts from the video subtitles and from the video frames and then we extract their lemmas\footnote{NLTK WordNet Lemmatizer \url{https://bit.ly/39XyX08}, retrieved Sep'19}. This resulted in an average of 36 entities (minimum 0, maximum 58) per video and an average of 175 labels (minimum 34, maximum 524) per video.

\paragraph{Key Concepts Selection}
It would be difficult to visualize in a word cloud all concepts extracted from video subtitles and frames, as they could reach values higher than 524. Thus, we chose to show only the \emph{key concepts} from the two sources:

\noindent \emph{a). Entities from Video Subtitles}: select all entities of type event, location, organization and people and exclude the other types. Upon analysis, this selection gives us the most optimal set of named entities, which most likely will not appear among the video frames labels. Furthermore, previous research \cite{de2017enriching,gligorov2011role} showed these types are the most useful for contextualizing information in videos. 

\noindent \emph{b). Labels from Video Frames}:
    \begin{inparaenum}[(1)]
        \item select the labels that appear in all three portions of the video, \emph{i.e.}, start, middle, end;
        \item among these, select the labels that have an occurrence higher than the median value in the video that is being analyzed.
    \end{inparaenum}

Subsequently, we have an average of 18 entities (minimum 0, maximum 44) and an average of 17 labels (minimum 0, maximum 46) per video. We used the union of these key concepts (average of 34, minimum 4 and maximum 73) to generate the explanations.

In Section \ref{sec:limitations} we discuss the limitations of our approach, namely the use of automated tools and the selection of key concepts.

\paragraph{Prominence of Key Concepts} 
We normalize the frequency of concepts in video subtitles and video frames by the maximum occurrence among each of them, to have prominence values from (0, 1] for both types of concepts. If a concept appears in both subtitles and frames, the word cloud shows the prominence from the subtitles.


\section{User Studies}
\label{sec:user_study}

We conducted two online user studies: \emph{User study 1} to find out which type of visual explanation is preferred to assess the representativeness of video summaries, and \emph{User Study 2} to understand how one (style of) visual explanation is used by participants to compare two summaries of the same video for their representativeness. 

\emph{Subjects:} We conducted the studies on the Amazon Mechanical Turk (AMT) platform, with master workers, with at least 98\% acceptance rate, more than 500 accepted HITs, and from English-speaking countries (US, UK, CA, AU). 

\subsection{User Study 1}
\label{subsec:user_study_1}

The first user study\footnote{Template: \url{https://github.com/oana-inel/FAIRView-VideoSummaryExplanations/blob/master/user_study1/user_study1_template.md}} aims to identify which of the visual explanations in Section \ref{sec:types_of_explanations} contribute the most to understanding how representative a video summary is, with regard to the original video. 

\paragraph{Materials} 
We used 20 videos and one of their summaries of 10 seconds (see Section \ref{sec:dataset_video_summaries}). We manually selected the videos in order to achieve a balanced and unbiased set of video summaries, \emph{i.e.}, which covers a large range of characteristics and represent well all four dimensions identified from the user preferences, \emph{i.e.}, \emph{semantic coverage}, \emph{semantic prominence}, \emph{quantity coverage} and \emph{distance}.
\begin{itemize}[noitemsep,nosep,topsep=0pt,leftmargin=*]
    \item \textbf{Many Prominent (MP)} 10 videos with many prominent concepts (high amount of concepts that appear throughout the video)
    \begin{itemize}
        \item \textbf{Many Prominent - Many Covered (MP-MC)} 5 summaries have high overlap on the prominent concepts 
        \item \textbf{Many Prominent - Few Covered (MP-FC)} 5 summaries have high overlap on the prominent concepts
    \end{itemize}
    \item \textbf{Few Prominent (FP)} 10 videos without many prominent concepts, but with similar prominence throughout the video
    \begin{itemize}
        \item \textbf{Few Prominent - Many Covered (FP-MC)} 5 summaries have high overlap on the prominent concepts 
        \item \textbf{Few Prominent - Few Covered (FP-FC)} 5 summaries have low overlap on the prominent concepts
    \end{itemize}
\end{itemize}

\paragraph{Independent Variables}
For the videos above, we generated four explanations with different dimensions of transparency, as described in Section \ref{sec:types_of_explanations}: \texttt{Summary WordCloud} (\emph{semantic coverage} \& \emph{semantic prominence}), \texttt{Overlap Fraction} (\emph{quantity coverage} \& \emph{distance}), \texttt{Overlap WordCloud} (\emph{semantic coverage}, \emph{semantic prominence} \& \emph{distance}) and \texttt{Combined WordCloud + Fraction} (all four).

\paragraph{Procedure}
Each video and summary was annotated by 20 participants and each judgment was remunerated with \$0.45. The procedure of the study follows the steps below:

\emph{1. Watch videos}: The participants first watch the full video. When the video finished, they are asked to watch the full summary of the video. The video summary is hidden from the participants unless the video has finished playing. Similarly, the questions of the study are not available unless the video summary has finished playing. 

\emph{2. Answer questions}: First, we informed participants about the goal of the study, \emph{i.e.}, to understand how we could support video archives in summarizing their content. 
Then, we asked them two questions, for which the answers were randomized each time. \textbf{(1)} We asked participants to inspect the four types of explanations (order was randomized each time) and choose the ones they find the most useful to understand how representative the video summary is for the video they watched. \textbf{(2)} We asked them to motivate their answer by choosing all the reasons that apply from a predefined list. 
Finally, we asked the participants to tell us which other visualization they find useful for assessing representativeness.  

\paragraph{Hypotheses}
In this user study we test the following hypotheses:
\Hypothesis{0} {The baseline explanation, \emph{Summary WordCloud}, has the least utility for understanding the representativeness of video summaries.} This explanation does not have any information about the original video, and does not contain any of the additional information identified in the user preferences. We expect it to have low utility.

\Hypothesis{1} {The dimension of \emph{distance} brings meaningful utility for understanding the representativeness of video summaries.}

\Hypothesis{2} {The dimensions of \emph{semantic coverage} and \emph{semantic prominence} have more utility than the dimension of \emph{quantity coverage} for understanding the representativeness of video summaries.}

\Hypothesis{3} {The combination of the four dimensions provides the best utility for understanding the representativeness of video summaries.}


\subsection{User Study 2}
\label{subsec:user_study_2}
The first study confirmed that the identified explanatory dimensions were helpful for assessing the representativeness of video summaries. However, we do not know how participants would use this information. Thus, in this study\footnote{Template: \url{https://github.com/oana-inel/FAIRView-VideoSummaryExplanations/blob/master/user_study2/user_study2_template.md}}, we compare two different video summaries of the same video and their \texttt{Combined WordCloud + Fraction} explanation (\emph{i.e.}, the explanation with the highest representativeness utility as resulted from \emph{User Study 1}). Since these summaries will be differently representative, this allows us to study in further depth how people use this kind of visual information.

\paragraph{Materials} 
We used 18 of the videos from the first study, their two summaries of 10 seconds, for consistency, and their visual explanations \emph{Combined WordCloud + Fraction}. We excluded two videos because explanations could not be generated for them (one video summary had no concept overlap with the key concepts of the original video - Section \ref{sec:methodology_of_explanation_generation}).

\paragraph{Procedure}
20 participants compared each pair of summaries and explanations and each comparison was remunerated with 0.45 USD. The procedure of the study follows the steps below:

\emph{1. Watch videos}: The participants first watch the video and then the two video summaries. The video summaries are hidden from the participants unless the video has finished playing and the questions of the study are not available unless the video summaries have finished playing. According to \textbf{U5}, the visual explanations are shown under each video summary, from the moment the video summaries are shown. The order of the two summaries is randomized.

\emph{2. Answer questions}: We provided the same context for the study, \emph{i.e.}, to understand how we could support video archives in summarizing their content. First, we asked the participants to use the explanations of the video summaries to decide how representative the two summaries (VS1 and VS2) are for the original video, by choosing one option among: ``VS1 is more representative than VS2'', ``VS2 is more representative than VS1'', ``VS1 and VS2 are equally representative'' or ``VS1 and VS2 are not representative''. We also asked them to motivate their answer. 

\paragraph{Hypothesis}
In this user study we test the following hypothesis:
\Hypothesis{4} {Users make use of the four dimensions identified as meaningful in Section \ref{sec:types_of_explanations} to assess the representativeness of video summaries.}


\section{Results}
\label{sec:results}

In this section we analyze and present the results of our user studies.

\paragraph{Analytical Method}
To test our hypotheses H0 to H3 we apply Cochran's Q test, a non-parametric statistical test which is similar to a Chi test, but can be applied when multiple categories are selected for the same item, as in our case. For all hypotheses we have the following null hypothesis: there is no difference in the percentage of participants that prefer each explanation, where the set of explanations varies for each hypothesis. To test hypothesis H4 we perform a qualitative analysis of the participants' comments. 

\subsection{User Study 1}
\label{subsec:results_user_study_1}

We report on the results of our first user study by testing the hypotheses presented in Section \ref{subsec:user_study_1}. We test the utility of the four dimensions, \emph{semantic coverage}, \emph{semantic prominence}, \emph{quantity coverage} and \emph{distance} to understand the video summaries representativeness. Each dimension characterizes one or more visual explanations.

\subsubsection{Participants}
We gathered 400 judgments from 69 participants. Each participant annotated around 6 videos, 31 participants annotated only one video, and 7 participants annotated all 20 videos. 

\subsubsection{Overview of Participants Preferences}
Explanation \texttt{Combined WordCloud + Fraction} received the most votes, being chosen in \np{194} out of \np{400} answers, and it is followed by explanation \texttt{Overlap WordCloud} with \np{106} out of \np{400} votes. The least votes are received by explanation \texttt{Overlap Fraction}, with not more than \np{68} votes and explanation \texttt{Summary WordCloud} with \np{87} votes. 

In Table \ref{tab:explanation_vs_prominence} we report the percentage of answers for each visual explanation on the four video categories used in our study (see Section \ref{subsec:user_study_1}). For every video category, explanation \texttt{Combined WordCloud + Fraction} is the most chosen as being useful for understanding how representative the video summary is for the original video. Furthermore, explanation \texttt{Overlap Fraction} has the least representativeness utility for three video categories, namely MP-MC, LP-MC and LP-FC. The exception are, however, the videos with many prominent concepts and few covered in the video summary, (MP-FC), where the least preferred explanation is instead \texttt{Summary WordCloud}, and then followed by \texttt{Overlap Fraction}.

Following, we test our four hypothesis (analysis in Table \ref{tab:statistical_analysis_study1}).

\begin{table*}[!h]
\caption{Results of the Cochran's Q test for each hypothesis in User Study 1, for all videos in our dataset (All Videos) and per video category. The visual explanations compared in each hypothesis are abbreviated: \texttt{Summary WordCloud} - SWC, \texttt{Overlap Fraction} - OF, \texttt{Overlap WordCloud} - OWC and \texttt{Combined WordCloud + Fraction} - CWC+F}
\label{tab:statistical_analysis_study1}
\resizebox{0.9\textwidth}{!}{
\begin{tabular}{cclcclclc}
 & H0 &  & \multicolumn{2}{c}{H1} &  & H2 &  & H3 \\ \cline{2-2} \cline{4-5} \cline{7-7} \cline{9-9} 
 & SWC vs. OF &  & OWC vs. SWC & OF vs. SWC &  & OF vs. OWC &  & CWC+F vs. all \\ \hline
All Videos  & \begin{tabular}[c]{@{}c@{}}Cochran's Q = 2.39\\ p = 0.12\end{tabular}          &  & \begin{tabular}[c]{@{}c@{}}Cochran's Q = 2.13\\ p = 0.14\end{tabular}         & \begin{tabular}[c]{@{}c@{}}Cochran's Q = 2.39\\ p = 0.12\end{tabular}          &  & \textbf{\begin{tabular}[c]{@{}c@{}}Cochran's Q = 9.75\\ p = 0.001\end{tabular}} &  & \textbf{\begin{tabular}[c]{@{}c@{}}Cochran's Q = 90.16\\ p = 2.02e$^{-19}$\end{tabular}} \\ \hline
MP-MC       & -           &  & \begin{tabular}[c]{@{}c@{}}Cochran's Q = 0.53\\ p = 0.46\end{tabular}         & \begin{tabular}[c]{@{}c@{}}Cochran's Q = 0.4\\ p = 0.52\end{tabular}           &  & \begin{tabular}[c]{@{}c@{}}Cochran's Q = 2.31\\ p = 0.12\end{tabular}           &  & \textbf{\begin{tabular}[c]{@{}c@{}}Cochran's Q = 17.89\\ p = 0.0004\end{tabular}}        \\ \hline
MP-FC       & \begin{tabular}[c]{@{}c@{}}Cochran's Q = 1.4\\ p = 0.23\end{tabular}           &  & \textbf{\begin{tabular}[c]{@{}c@{}}Cochran's Q = 6.4\\ p = 0.01\end{tabular}} & \begin{tabular}[c]{@{}c@{}}Cochran's Q = 1.4\\ p = 0.23\end{tabular}           &  & \begin{tabular}[c]{@{}c@{}}Cochran's Q = 1.8\\ p = 0.17\end{tabular}            &  & \textbf{\begin{tabular}[c]{@{}c@{}}Cochran's Q = 24.87\\ p = 1.64e$^{-05}$\end{tabular}} \\ \hline
FP-MC      & - &  & \begin{tabular}[c]{@{}c@{}}Cochran's Q = 2.18\\ p = 0.13\end{tabular}         & \textbf{\begin{tabular}[c]{@{}c@{}}Cochran's Q = 4.45\\ p = 0.03\end{tabular}} &  & \begin{tabular}[c]{@{}c@{}}Cochran's Q = 0.75\\ p = 0.38\end{tabular}           &  & \textbf{\begin{tabular}[c]{@{}c@{}}Cochran's Q = 27.57\\ p = 4.45e$^{-06}$\end{tabular}} \\ \hline
FP-FC     & -       &  & \begin{tabular}[c]{@{}c@{}}Cochran's Q = 1.08\\ p = 0.29\end{tabular}         & \begin{tabular}[c]{@{}c@{}}Cochran's Q = 2.0\\ p = 0.15\end{tabular}           &  & \textbf{\begin{tabular}[c]{@{}c@{}}Cochran's Q = 6.42\\ p = 0.01\end{tabular}}  &  & \textbf{\begin{tabular}[c]{@{}c@{}}Cochran's Q = 28.12\\ p = 3.42e$^{-06}$\end{tabular}} \\ \hline \hline
\end{tabular}
}
\end{table*}

\paragraph{H0: The baseline explanation, \texttt{Summary WordCloud}, has the least utility for understanding the representativeness of video summaries.} As we saw in Table \ref{tab:explanation_vs_prominence}, the visual explanation \texttt{Summary WordCloud} did not received the least amount of votes overall. There was an exception for videos with many prominent concepts and few covered in the summary, MP-FC. However, in Table \ref{tab:statistical_analysis_study1} we see according to Cochran's Q test (p > 0.05) that the difference between the baseline and the \texttt{Overlap Fraction} explanation (\emph{i.e.}, the following preferred explanation) was not statistically significant. 
Overall, we reject our \textbf{hypothesis H0}. We did not find support that this explanation has the lowest utility overall.


\paragraph{H1: The dimension of distance brings meaningful utility for understanding the representativeness of video summaries, compared to the baseline.} The visual explanations \texttt{Overlap Fraction} and \texttt{Overlap WordCloud} are characterized by the dimension of \emph{distance}. Thus, we analyze whether any of these visualizations is preferred over the baseline, \emph{i.e.}, \texttt{Summary WordCloud}. From Table \ref{tab:explanation_vs_prominence} and Table \ref{tab:statistical_analysis_study1}, we know that actually, the baseline explanation is statistically preferred over the \texttt{Overlap Fraction} explanation for FP-MC videos, and not vice versa. Further, we also observe that explanation \texttt{Overlap WordCloud} for videos with many prominent concepts and few covered in the summary is preferred over the baseline. \textbf{H1} is accepted for only one type of video, MP-FC, for \texttt{Overlap WordCloud}.

\paragraph{H2: The dimensions of semantic coverage and prominence have more utility than the dimension of quantity coverage for understanding the representativeness of video summaries.} We test here whether the visual explanation \texttt{Overlap WordCloud} is more preferred than the visual explanation \texttt{Overlap Fraction}. In Table \ref{tab:explanation_vs_prominence} we observe that \texttt{Overlap WordCloud} receives more votes than \texttt{Overlap Fraction} on all video types. According to Table \ref{tab:statistical_analysis_study1}, our hypothesis \textbf{H2} is accepted only for videos with few prominent concepts and few covered concepts in the video summary, where the preference is statistically significant.  

\paragraph{H3: The combination of the four dimensions provides the best utility for understanding the representativeness of video summaries.} The \texttt{Combined WordCloud + Fraction} visual explanation incorporates all four dimensions that are considered relevant for understanding representativeness of video summaries. As observed in Table \ref{tab:statistical_analysis_study1}, we reject the null hypothesis, for all videos and for all video categories, which means that at least one visual explanation is significantly preferred compared to the others. This, in addition to the statistics in Table \ref{tab:explanation_vs_prominence} show that our hypothesis \textbf{H3} is accepted, \emph{i.e.}, explanation \textbf{\emph{Combined WordCloud + Fraction}} is always preferred.

\begin{table}[h]
\caption{Percentage of answers (smallest in bold) containing each visual explanation on the four video categories used in our study (Many Prominent - Many Covered (MP-MC),  Many Prominent - Few Covered (MP-FC), Few Prominent - Many Covered (FP-MC), Few Prominent - Few Covered (FP-FC).}
\label{tab:explanation_vs_prominence}
\resizebox{0.9\columnwidth}{!}{
\begin{tabular}{lcccc}
      & \begin{tabular}[c]{@{}c@{}}Summary \\ WordCloud\end{tabular}  & \begin{tabular}[c]{@{}c@{}}Overlap \\Fraction\end{tabular} & \begin{tabular}[c]{@{}c@{}}Overlap \\WordCloud\end{tabular} & \begin{tabular}[c]{@{}c@{}}Combined \\WordCloud \& \\Fraction\end{tabular}  \\ \hline \hline
MP-MC & 22\%   & \textbf{18\%}  & 27\%  & 46\%   \\
MP-FC & \textbf{15\%}   & 22\%  & 31\%  & 49\%   \\ 
FP-MC & 29\%  & \textbf{15\%}  & 20\%  & 50\%   \\
FP-FC & 21\%   & \textbf{13\%}  & 28\%  & 49\%  \\ \hline
\end{tabular}
}
\end{table}

\subsubsection{Visual Explanation Motivation}
The study participants motivate their visual explanation preferences, by mentioning that they find it useful, in equal proportion of 34\%, to see \emph{(1)} which concepts from the original video were covered in the summary and \emph{(2)} which are the prominent concepts of the original video. Conversely, the least chosen motivations were \emph{(1)} ``It was useful to see the concepts of the video summary'' in proportion of 25\% and \emph{(2)} ``It was useful to see the similarities between summary and original video'' in proportion of 27\%. Therefore, the choice of the motivations further emphasize the preference of the participants towards the visual explanation \textbf{\emph{Combined WordCloud + Fraction}}.

\subsubsection{Additional Visual Explanations}
In general, participants mentioned that the four visual explanations provide a comprehensive understanding of the representativeness of video summaries. 
Among the comments received, we summarize the most informative ones with regard to additional visual explanations: \emph{(1)} a word count comparison; \emph{(2)} a breakdown on the main topics and the percentage of the topic in the original video and in the summary; \emph{(3)} a summary of the percentage of time devoted to each speaker in the original video and in the summary; \emph{(4)} the location of the topics in the original video, \emph{i.e.}, beginning, middle, end; \emph{(5)} footnotes attached to the words in the summary to go into a more detailed summary when clicking on it. Furthermore, participants also suggested the addition of ``bar chart'' and ``Venn diagram'' for showing which concepts are covered by the video summary and what is their percentage. These suggestions are further discussed in Section \ref{sec:conclusions}, as future work.

\subsection{User Study 2}
We now report on the results of \emph{user study 2}, where participants were asked to assess the representativeness of two video summaries and their \texttt{Combined WordCloud + Fraction} explanation. We perform a qualitative analysis on the participants' comments. 

\subsubsection{Participants} We gathered 360 judgments from 76 participants. Each participant annotated around 5 videos, 7 participants annotated all videos and 36 participants annotated only one video.

\subsubsection{Overview of summary representativeness}
In general, participants consider the two video summaries equally representative (126 votes out of 360). 46 times none of the summaries was considered representative, while in general, the two summaries receive equal amount of votes (97 vs. 91). 

\emph{H4: Users make use of all four dimensions of transparency to compare the representativeness of video summaries.} Following, we perform a qualitative analysis of the participants' comments.

When participants consider none of the summaries as representative, they motivate their choice by stating that the focus was not on the main topics, thus by making reference to the \emph{semantic coverage} and \emph{distance}, \emph{e.g.}, \emph{``Neither captures the main points that the video is about.''}, \emph{``Neither ... deals with the main theme of the video.''}, \emph{``Both have small variations but do not show enough of the main topics to get an understanding of the events.''}. 

When the video summaries are very similar and they seem to cover the same topics, participants tend to reference the \emph{quantity coverage}: \emph{``... the emphasis is different, but the percentages of concepts in the video summaries are the same''}, \emph{``... same percentage found in the original''} or simply the similarities between the two summaries: \emph{``same information each.''}, \emph{``both are good, good information and good visual images, good context''}. Participants also comment on the fact that the two summaries cover the same key concepts of the original video, such as, \emph{``both seemed to hit the same key points from he main video''} or \emph{``both cover some key points, they are both good.''}. Since only a small number of participants said none of the video summaries is representative, namely 12.78\%, we were not able to understand whether video summary quality can affect user comprehension.

When choosing one of the two summaries as being more representative, two reasons are emphasized. On the one hand, participants check the concepts coverage in percentage, thus looking into \emph{quantity coverage}: \emph{``had more of the topics of the video covered''}, \emph{``covers 10\% more of the video concepts''}, \emph{``more video contents are covered-6\% more''}. On the other hand, they verify the \emph{semantic coverage} and \emph{prominence} of the topic mentioned: \emph{``The words and summary shown in this image are more prevalent to the original clip.''}, \emph{``While the percent of video concepts found is the same, Summary 1 expresses the importance of concepts found in the word cloud.''}, \emph{``Summary 1 covers more of what is in the original video and the word cloud generated highlights more of the hurricane problems discussed.}. It is also interesting to notice in the examples above that participants mention how each component of the visualization helped them decide on the representativeness of the video summaries.

Our \textbf{hypothesis H4} is accepted, as we found evidence in participants' comments that all four transparency dimensions contribute to the decision of choosing the most representative video summary.


\section{Discussion}
\label{sec:discussion}

In general, we observe that participants \emph{prefer to see which are the most prominent concepts in the original video that are covered or not covered in the video summary}. This observation is also supported by current literature on explanations \cite{lim2009and}, which concludes that people need explanations that contain both arguments and counter arguments. Additionally, participants prefer \emph{the most complete visual explanations} that \emph{semantically characterizes (word cloud with semantic and prominence overlap), and quantifies (donut with percentage overlap) the overlap between the video summary and the original video, which reflecting the differences between the two, \emph{distance}}. Therefore, all four dimensions of transparency identified in our user preferences elicitation survey are meaningful. 

Interestingly, the baseline explanation, \texttt{Summary WordCloud} is the least preferred for just one type of videos, namely videos with many prominent concepts and few covered in the video summary and the second least preferred for the other video categories. Conversely, the visual explanation \texttt{Overlap Fraction} is more preferred on videos with many prominent concepts and few covered in the video summary and the least preferred on all the other categories. This indicates that, in general, the \emph{quantity coverage} dimension has the least representativeness utility when used alone. Furthermore, participants have a strong tendency towards understanding the \emph{semantic coverage} at the level of the video summary, and even more prominently at the level of both video summary and original video.    

When the visual explanations are used in practice to assess the representativeness of video summaries, we observe that all four dimensions of transparency shape the decision of the participants. Although the dimension of \emph{quantity overlap} expressed in the \texttt{Overlap Fraction} has the least utility in assessing representativeness, participants still use it when the two video summaries are quite similar to help them decide whether they are equally representative or one is more representative than the other. The key concepts of the video, expressing \emph{semantic coverage}, \emph{semantic prominence} as well as \emph{distance} seem to be the main aspect of assessing representativeness of video summaries. This emphasizes the need to provide users with highly meaningful video concepts.

\subsection{Limitations}
\label{sec:limitations}

We identified several limitations of this work, regarding \emph{(1)} the visual explanation generation approach, \emph{(2)} the experimental setup and \emph{(3)} the types of video summaries used in our experiments.

\emph{Visual explanation generation approach}: consists of automated components. First, we rely on a single annotation tool and its ability to extract meaningful concepts from videos. Some concepts might be misleading or too general (\emph{e.g.}, ``event'' is quite often extracted from the video stream). Second, we extract entities from automatically generated video subtitles. Thus, entities that are not well recognized by speech-to-text may not be correctly identified. Third, there is a gap between machines and humans \cite{gligorov2011role}, meaning that people may describe a video with different concepts than machines. Thus, the key concepts we identify may not be the most representative for the video. The results of our second user study also emphasized the need to visualize the key concepts, as many participants base their decision on them. Nonetheless, our key concepts still convey a lot of information, as showed in our experiments. The visual explanation generation approach is generalizable for any video summary length. However, longer videos could have an impact, since more concepts need to be included in the visual explanations and people would need to take longer to understand the representativeness of the summary.

\emph{Experimental setup}: does not consider the overall quality of the video summaries. We are not aware if video summary representativeness can be drastically affected by the overall quality of the video summaries. This issue is not the focus of our work, however, we try to mitigate it by comparing video summaries of equal length and generated by the same video summarizer, which contain at least some of the key concepts of the original video. Another limitation regarding the experimental setup consists in the limited number of visual explanations explored. However, the visual explanations explored were the most suitable to incorporate the user preferences and the four dimensions of transparency. 

\emph{Different types of summaries}: are not considered. We did not analyze the effects of \emph{(1)} the \emph{video summary duration} (we only used summaries of 10 seconds) and of \emph{(2)} different types of summarization (we only used keyshot-based video summaries).


\section{Conclusions and Future Work}
\label{sec:conclusions}

In this paper we proposed a set of four visual explanations to guide users to understand how representative a video summary is with regard to the original video. The proposed visual explanations were generated based on \emph{user preferences} gathered from target users such as media scholars, archivist and collection owners. The explanations provide overviews of the summary, differences and overlaps between the original video and the summary, among others. Four \emph{dimensions of transparency} are covered by our proposed explanations: \emph{semantic coverage}, \emph{semantic prominence}, \emph{quantity coverage} and \emph{distance}. Our results suggest that people prefer the highest level of transparency (combining all four dimensions) when assessing the representativeness of a summary with respect to the original video. More precisely, they prefer to see the exact concepts that overlap and differ between the original video and the summary, as well as the percentage of concepts that is covered by the summary. Moreover, when comparing two video summaries of the same video, each dimension of transparency contributes to the choice of the most representative one.

In future work we plan to look into different approaches for identifying the key concepts in videos, by means of human-in-the-loop approaches for video enrichment, that also mitigate the gap between humans and machines in this task. Moreover, we plan to extend this work by studying \emph{(1)} the effect of the video summary duration and other video summary types and \emph{(2)} the suitability of other visual representations, as suggested in the first user study. 

\section{Acknowledgments}
The authors thank the anonymous reviewers for their valuable comments and the user studies participants for their participation and helpful insights. The authors also thank the Netherlands Institute for Sound and Vision and The Center for Research and Technology for their support and input in this research. This work was supported by the Google DNI Fund and Google Research credits.  
\bibliographystyle{ACM-Reference-Format}
\bibliography{main_iui2020}

\end{document}